\magnification=1000
\def\pspicture #1 by #2 (#3){
  \vbox to #2{
    \hrule width #1 height 0pt depth 0pt
    \vfill
    \includegraphics{#3} 
    }
  }
\def\scaledpspicture #1 by #2 (#3 scaled #4){{
  \dimen0=#1 \dimen1=#2
  \divide\dimen0 by 1000 \multiply\dimen0 by #4
  \divide\dimen1 by 1000 \multiply\dimen1 by #4
  \pspicture \dimen0 by \dimen1 (#3)}
  }
\def\Molla{\scaledpspicture 424pt by 141pt  (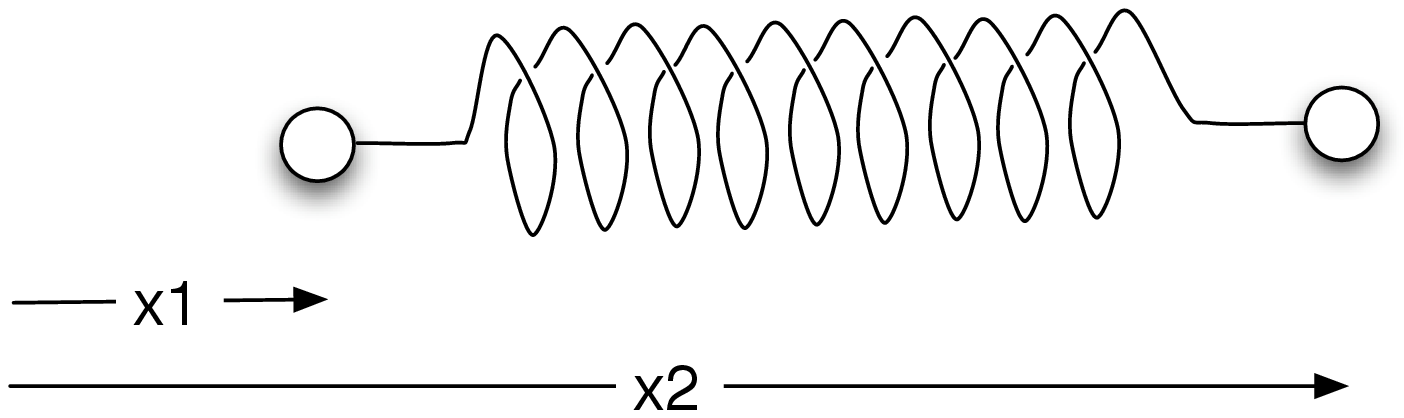 hscale=30 vscale=30 scaled 300)}
%
%
%
\def\SansSerif{cmss}
=\SansSerif10  at 14pt
=\SansSerif10 
=\SansSerif9
=\SansSerif8 
=\SansSerif7
=\SansSerif6
=\SansSerif5

\def\Serif{cmr}
=\Serif10
=\Serif9
=\Serif8
=\Serif7
=\Serif6
=\Serif5

\font\tencmti=cmti10

\font\eightcmti=cmti8


\font\tencmsl=cmsl10

\font\eightcmsl=cmsl8


\font\tencmbx=cmbx10

\font\eightcmbx=cmbx8
\font\sevencmbx=cmbx7
\font\sixcmbx=cmbx6
\font\fivecmbx=cmbx5

\font\tencmmi=cmmi10

\font\sevencmmi=cmmi7

\font\fivecmmi=cmmi5

\font\tencmsy=cmsy10

\font\eightcmsy=cmsy8
\font\sevencmsy=cmsy7

\font\fivecmsy=cmsy5

\font\tencmex=cmex10

\font\sevencmex=cmex7

\font\fivecmex=cmex7  at 5pt

\font\tencmtt=cmtt10

\font\tencmmib=cmmib10

\font\eightcmmib=cmmib8

\font\teneufm=eufm10

\font\eighteufm=eufm8

\font\tenmsbm=msbm10

\font\eightmsbm=msbm7 at 8pt


 \def\sec{\tencmbx}        
 \def\ssec{\tencmbx}          

\def\AbsStyle{\normalbaselineskip=10pt%
\def\nt{\eightcmr}%
\def\rm{\fam0\eightcmr}%
\textfont0=\tencmr   \scriptfont0=\sevencmr   \scriptscriptfont0=\fivecmr%
\textfont1=\tencmmi  \scriptfont1=\sevencmmi  \scriptscriptfont1=\fivecmmi%
\textfont2=\tencmsy  \scriptfont2=\sevencmsy  \scriptscriptfont2=\fivecmsy%
\textfont3=\tencmex  \scriptfont3=\sevencmex  \scriptscriptfont3=\fivecmex%
\textfont\itfam=\eightcmti \def\it{\fam\itfam\eightcmti}%
\textfont\slfam=\eightcmsl \def\sl{\fam\slfam\eightcmsl}%
\textfont\ttfam=\tencmtt \def\tt{\fam\ttfam\tencmtt}%
\textfont\bffam=\eightcmbx\def\bf{\fam\bffam\eightcmbx}%
\scriptfont\bffam=\sixcmbx \scriptscriptfont\bffam=\fivecmbx%
\def\gothic{\eighteufm}%
\def\cal{\eightcmsy}%
\def\bg{\eightcmmib}%
\def\Bbb{\eightmsbm}%
\def\LieFont{\eightcmti}%
\nt\normalbaselines\baselineskip=12pt%
}

\def\DimStyle{\normalbaselineskip=12pt%
\def\nt{\tencmsl}%
\def\rm{\fam0\tencmr}%
\textfont0=\tencmr  \scriptfont0=\sevencmr  \scriptscriptfont0=\fivecmr%
\textfont1=\tencmmi  \scriptfont1=\sevencmmi  \scriptscriptfont1=\fivecmmi%
\textfont2=\tencmsy  \scriptfont2=\sevencmsy  \scriptscriptfont2=\fivecmsy%
\textfont3=\tencmex  \scriptfont3=\sevencmex  \scriptscriptfont3=\fivecmex%
\textfont\itfam=\tencmti \def\it{\fam\itfam\tencmti}%
\textfont\slfam=\tencmr  \def\sl{\fam\slfam\tencmr}%
\textfont\ttfam=\tencmtt \def\tt{\fam\ttfam\tencmtt}%
\textfont\bffam=\tencmbx \def\bf{\fam\bffam\tencmbx}%
\scriptfont\bffam=\sevencmbx \scriptscriptfont\bffam=\fivecmbx%
\def\gothic{\teneufm}%
\def\cal{\tencmsy}%
\def\bg{\tencmmib}%
\def\Bbb{\tenmsbm}%
\def\LieFont{\tencmti}%
\nt\normalbaselines\baselineskip=15pt%
}

\def\ClaimStyle{\normalbaselineskip=12pt%
\def\nt{\tencmr}%
\def\rm{\fam0\tencmr}%
\textfont0=\tencmr  \scriptfont0=\sevencmr  \scriptscriptfont0=\fivecmr%
\textfont1=\tencmmi  \scriptfont1=\sevencmmi  \scriptscriptfont1=\fivecmmi%
\textfont2=\tencmsy  \scriptfont2=\sevencmsy  \scriptscriptfont2=\fivecmsy%
\textfont3=\tencmex  \scriptfont3=\sevencmex  \scriptscriptfont3=\fivecmex%
\textfont\itfam=\tencmti \def\it{\fam\itfam\tencmti}%
\textfont\slfam=\tencmsl \def\sl{\fam\slfam\tencmsl}%
\textfont\ttfam=\tencmtt \def\tt{\fam\ttfam\tencmtt}%
\textfont\bffam=\tencmbx \def\bf{\fam\bffam\tencmbx}%
\scriptfont\bffam=\sevencmbx \scriptscriptfont\bffam=\fivecmbx%
\def\gothic{\teneufm}%
\def\cal{\tencmsy}%
\def\bg{\tencmmib}%
\def\Bbb{\tenmsbm}%
\def\LieFont{\tencmti}%
\nt\normalbaselines\baselineskip=15pt%
}

\def\NormalStyle{\normalbaselineskip=12pt%
\def\nt{\tencmr}%
\def\rm{\fam0\tencmr}%
\textfont0=\tencmr  \scriptfont0=\sevencmr  \scriptscriptfont0=\fivecmr%
\textfont1=\tencmmi  \scriptfont1=\sevencmmi  \scriptscriptfont1=\fivecmmi%
\textfont2=\tencmsy  \scriptfont2=\sevencmsy  \scriptscriptfont2=\fivecmsy%
\textfont3=\tencmex  \scriptfont3=\sevencmex  \scriptscriptfont3=\fivecmex%
\textfont\itfam=\tencmti \def\it{\fam\itfam\tencmti}%
\textfont\slfam=\tencmsl \def\sl{\fam\slfam\tencmsl}%
\textfont\ttfam=\tencmtt \def\tt{\fam\ttfam\tencmtt}%
\textfont\bffam=\tencmbx \def\bf{\fam\bffam\tencmbx}%
\scriptfont\bffam=\sevencmbx \scriptscriptfont\bffam=\fivecmbx%
\def\gothic{\teneufm}%
\def\cal{\tencmsy}%
\def\bg{\tencmmib}%
\def\Bbb{\tenmsbm}%
\def\LieFont{\tencmti}%
\nt\normalbaselines\baselineskip=15pt%
}

\NormalStyle

\def\ni{\noindent}
\def\ss{\vskip 5pt}
\def\ms{\vskip 10pt}
\def\bs{\vskip 15pt}

\def\nl{\goodbreak\ni}
\long\def\Title#1{\ni{\fourteenss #1}}
\long\def\Abstract#1{\vskip30pt\ni{\leftskip25pt\rightskip10pt{\AbsStyle \bf Abstract:\ }%
{\AbsStyle #1}\par}\vskip20pt}
\long\def\Author#1{\bs\ni{\tencmss #1}\par}
\long\def\Address#1{\ss\ni{\tencmss #1}\par}
%
%
\newcount\PART		          %
\newcount\CHAPTER		          %
\newcount\SECTION		          %
\newcount\SUBSECTION		       %
\newcount\FNUMBER            
\newdimen\TOBOTTOM
\newdimen\LIMIT

\SECTION=0		          
\SUBSECTION=0		       
\FNUMBER=0		          

\def\LastSection{Undefined}
\def\LastSubSection{Undefined}
\def\LastClaim{Undefined}
\def\Last{Undefined}
\def\SectionLabel{\the\SECTION.}
\long\def\NewSection#1{\global\advance\SECTION by 1%
         \bs\ni{\sec  \SectionLabel\ #1}\ss%
         \SUBSECTION=0\FNUMBER=0%
         \gdef\Left{#1}%
         \global\edef\Last{\SectionLabel}%
         \global\edef\LastSection{\SectionLabel}%
         \global\edef\LastSubSection{undefined}%
         \global\edef\LastClaim{undefined}}
\def\SubSectionLabel{\ifnum\SECTION>0 \the\SECTION.\fi\the\SUBSECTION.}
\long\def\NewSubSection#1{\global\advance\SUBSECTION by 1%
         \ms\ni{\ssec #1}\ss%
         \global\edef\Last{\SubSectionLabel}%
         \global\edef\LastSubSection{\SubSectionLabel}}

\def\fopen{(}\def\fclose{)}                               

\def\ClaimLabel{\fopen\ifnum\CHAPTER>0 \the\CHAPTER.\fi%
      \ifnum\SECTION>0 \the\SECTION.\fi%
      \the\FNUMBER\fclose}

\def\NewClaim{\global\advance\FNUMBER by 1%
    \ClaimLabel%
    \global\edef\LastClaim{\ClaimLabel}%
    \global\edef\Last{\ClaimLabel}}


\def\fn{\eqno{\NewClaim}}                         
\def\fl#1{\fn\global\edef#1{\ClaimLabel}}         


%
%
%
%

\def\ni{\noindent}
\def\ss{\vskip 5pt}
\def\ms{\vskip 10pt}

\def\noex{\noexpand}

\def\refs{}
\def\empty{\#}
\def\BibNumber{}
\def\BibTitle{}

\newcount\BNUM
\BNUM=0

\def\bib#1#2{\gdef#1{\global\def\BibNumber{\empty}\global\def\BibTitle{#2}}}

\def\ref#1{#1
\if\BibNumber\empty \global\advance\BNUM 1
\message{reference[\BibNumber]}\message{}
\global\edef\refs{\refs \ss\ni[\the\BNUM]\ \BibTitle}
\global\edef#1{\noex\global\noex\edef\noex\BibNumber{[\the\BNUM]}
 \noex\global\noex\edef\noex\BibTitle{\BibTitle}}
{\bf [\the\BNUM]}
\else
{\bf \BibNumber}
\fi}

\def\Biblio{{\refs}}

\def\al{\alpha}
\def\be{\beta}
\def\de{\delta}

\def\ep{\epsilon}

\def\la{\lambda}

\def\om{\omega}
\def\si{\sigma}

\def\De{\Delta}
\def\Ga{\Gamma}

 \def\one{{\hbox{\Bbb I}}}
 
 \def\R{{\hbox{\Bbb R}}}

 \def\E{{\hbox{\Bbb E}}}

 \def\F{{\hbox{\Bbb F}}}
 
 \def\R{{\hbox{\Bbb R}}}

\def\calL{{\hbox{\cal L}}}
\def\calE{{\hbox{\cal E}}}
\def\calB{{\hbox{\cal B}}}
\def\calU{{\hbox{\cal U}}}
\def\calW{{\hbox{\cal W}}}
\def\calQ{{\hbox{\cal Q}}}

\def\Div{{\hbox{Div}}}
\def\SO{{\hbox{SO}}}
\def\GL{{\hbox{GL}}}
\def\Lie{{\hbox{\LieFont \$}}}

\def\QDE{\hfill QDE}
\def\del{\partial}
\def\na{\nabla}
\def\d{\hbox{d}}
\def\arr{\rightarrow}

 \def\,{\mskip\thinmuskip}
 \def\!{\mskip-\thinmuskip}
 \def\>{\mskip\medmuskip}
 \def\;{\mskip\thickmuskip}


\bib{\Raiteri}{
M. Ferraris, M. Francaviglia, M. Raiteri, 
Classical Quantum Gravity 20(18) (2003),  4043--4066}

\bib{\Lovelock}{
G. Allemandi, M. Francaviglia, M.Raiteri, 
Classical Quantum Gravity 20 (2003), no. 23, 5103--5120. }

\bib{\Wald}{
V. Iyer and R. Wald, 
Phys. Rev. D {\bf 50},  1994, 846
}

\bib{\Silva}{
B. Julia, S. Silva, Class. Quantum Grav. 17, (2000), 4733 (gr-qc\/0005127); \goodbreak
S. Silva, Nucl. Phys. B 558, (1999), 391 (hep-th\/9809109).}

\bib{\NostroCS}{
A. Borowiec, M. Ferraris,M. Francaviglia,
 J. Phys. A 36 (2003), no. 10, 2589--2598.}

\bib{\Lagrangian}{
M. Ferraris, M. Francaviglia,
in: {\it Mechanics, Analysis and Geometry: 200 Years after Lagrange},
Editor: M. Francaviglia, Elsevier Science Publishers B.V., 1991}

\bib{\Remarks}{
L.ÊFatibene, M. Ferraris, M.Francaviglia, M. Raiteri, 
Ann. Physics 275(1) (1999), 27--53.}

\bib{\Libro}{
L. Fatibene, M. Francaviglia,
{\it  Natural and gauge natural formalism for classical field theories. A geometric perspective including spinors and gauge theories.}
 Kluwer Academic Publishers, Dordrecht, 2003. xxii+365 pp. ISBN: 1-4020-1703-0}

\bib{\Trautman}{
A.\ Trautman,
  in: {\it Papers in honour of J. L. Synge}, Clarenden Press, Oxford, 1972 UK,  85
}

\bib{\Cavalese}{
M. Ferraris and M. Francaviglia, 
in: {\it 8th Italian Conference on General Relativity and Gravitational Physics}, Cavalese (Trento), August 30 --
September 3, World Scientific, Singapore, 1988, 183
}

\bib{\Spencer}{
H. Goldschmidt, D. Spencer,
    J. Differential Geom. {\bf 13}(4) (1978), 455
}

\bib{\BTZ}{
L.ÊFatibene, M. Ferraris, M. Francaviglia, M. Raiteri,
 Phys. Rev. D (3) 60 (1999), no. 12, 124012, 7 pp.;
 L.ÊFatibene, M. Ferraris, M. Francaviglia, M. Raiteri,
   Phys. Rev. D (3) 60 (1999), no. 12, 124013, 10 pp.
 }

\bib{\Taub}{
L. Fatibene, M. Ferraris, M. Francaviglia, M. Raiteri, 
  Ann. Physics 284(2) (2000), 197--214.}

\bib{\HDTaub}{
R. Clarkson, L. Fatibene, R.B. Mann, 
 Nuclear Phys. B 652(1-3) (2003),  348--382.}

\bib{\Kats}{
J.\ Katz, 
Class.\ Quantum Grav., {\bf 2}, 1985, 423}

\bib{\ADMCovariante}{
M.\ Ferraris and M.\ Francaviglia, 
Atti Sem. Mat. Univ. Modena, {\bf 37}, 1989, 61;\goodbreak
M.\ Ferraris, M.\ Francaviglia and I.\ Sinicco, 
Il Nuovo Cimento, {\bf 107B},(11), 1992, 1303;\goodbreak
M.\ Ferraris and M.\ Francaviglia, 
Gen.\ Rel.\ Grav., {\bf 22}, (9), 1990.}

\bib{\Rovelli}{
Carlo Rovelli,
{\it Quantum Gravity},
Cambridge University Press, (2004) ISBN: 0521837332}

\bib{\Saunders}{
D. J. Saunders,  {\it The geometry of jet bundles},
 London Mathematical Society Lecture Note Series, 142. 
 Cambridge University Press, Cambridge, 1989.}

\bib{\Poincare}{
\'E. Cartan,
Hermann, Paris, 1922; reprint, 1958;\goodbreak
Th. H. J. Lepage, 
Bull. Acad. Roy. Belg. Cl. Sci. (5) 22 (1936), 716--729;\goodbreak
G.Giachetta, L. Mangiarotti,G. Sardanashvily,
 J. Phys. A 32 (1999), no. 38, 6629--6642. 
 }

\bib{\Altro}{
I.M. Anderson, C.G. Torre, Phys. Rev. Lett. 77 (1996) 4109 (hepÐ th/9608008); \goodbreak
C.G. Torre, hepÐth/9706092, Lectures given at 2nd Mexican School on Gravitation and Mathematical Physics, Tlaxcala, Mexico (1996). }

\bib{\Goldstein}{
H. Goldstein. {\it Classical Mechanics},
Addison Wesley, 1980.}

\parindent=10pt
\parskip=2pt
\baselineskip=12pt
\Title{Augmented Variational Principles and Relative Conservation Laws in Classical Field Theory}

\Author{L.Fatibene, M.Ferraris, M.Francaviglia}

\Address{Dipartimento di Matematica\nl 
Universit\`a degli Studi di Torino\nl
via Carlo Alberto 10\nl
10123 Torino\nl
ITALY}

\Abstract{%
Augmented variational principles are introduced in order to provide a definition of {\it relative conservation laws}.
As it is physically reasonable, relative conservation laws define in turn {\it relative conserved quantities}
which measure, for example, how much energy is needed in a field theory to go from one configuration (called the {\it reference} or {\it vacuum}) to another configuration (the {\it physical state} of the system).
The general prescription we describe solves in a covariant way the well known observer dependence of conserved quantities.
The solution found is deeply related to the divergence ambiguity of the Lagrangian and to various formalisms recently appeared in literature to deal with the variation of conserved quantities (of which this is a formal integration).
A number of examples relevant to fundamental Physics are considered in detail, starting from classical Mechanics.
}

\NewSection{Introduction}

\ni  
Conserved quantities are not absolute nor covariant in Physics.
Even in Mechanics the {\it kinetic energy} of a free particle is defined just after a rest frame has been fixed.
The same particle seen by an observer in uniform motion along a straight line has a different energy.

It can be trivially remarked that only differences of energy appear in classical Physics. 
What is sometimes called {\it the} energy of a system is just {\it one} energy relative to some choice (e.g., observer frame, another state of the system, the vacuum). 
Some of these choices are canonical (e.g. the vacuum of a Klein-Gordon  theory) some are not (e.g. the observer frame).  
This ambiguity in the definition of energy adds up to the usual (independent) ambiguity with respect to control modes which allow one to define different {\it ``thermodynamical''}
energies (e.g. internal energy, free energy, etc\dots). 

In relativistic field theories observers are often identified with coordinate systems; this argument  is sometimes used as a motivation in favor of coordinate dependent prescriptions for energy (also called pseudo-tensors prescriptions). 
We shall prove below that one can always define intrinsic and covariant quantities to represent a conserved quantity.
Hence pseudo-tensors are not physically necessary, but just a choice (and, unfortunately, a usually bad choice). 

Motivated by these naive considerations, we are going here to develop a framework for {\it relative} conserved quantities which represent, e.g., the amount of energy needed to {\it push} a system from one specific configuration to another. 
Such quantities (also in classical Physics) are independent of the observer (check by computing the energy needed to double the amplitude of a harmonic oscillator, both on the ground and on a train; if needed, see Appendix A) and are endowed with an intrinsic meaning (as testified by a habit of considering energy as a kind of price one has to {\it pay} to produce a definite effect). 
If we could pay less energy just by changing rest frame---which in principle can be done with zero energy since a classical observer can be as light as we wish---it would be beautiful but pretty unrealistic!

The new universal formalism we propose applies, reproducing usual results,  also to the cases in which a canonical choice is already {\it a priori} available for the vacuum state. It is important to notice that a canonical choice of a vacuum state is not always available.
In a field theory with configurations in a vector bundle  there is of course a canonical zero section which can be 
selected as a preferred vacuum state. 
However, in gauge theories the gauge fields live in an affine bundle (the bundles of connections of a fixed principal bundle $P$), where in general  there is no preferred section (since the choice $\bar A^a_\mu=0$ depends
on the trivialization, i.e. it is not gauge covariant and in most cases not even a global one).
Nevertheless (and luckily enough) there exists in gauge theories a class of connections which have zero curvature (the curvatures of principal connections live in fact in a vector bundle, so that {\it zero curvature} is not ambiguous). 
One can then select one of these flat connections as a vacuum state. 
Such a  choice is not unique but owing to gauge covariance the results do not depend on the representative chosen.

Finally in (all purely metric formalisms for) General Relativity (GR) configurations are non-degenerate metrics of some fixed signature. Once again one could select a Riemann-flat representative for the vacuum state (all solutions without matter are Ricci-flat).
Unfortunately this option is not available when, e.g., a cosmological constant is allowed (Minkowski is not a critical section and (anti)de-Sitter spaces are not flat).

In GR a solution to this problem is known; it amounts to leave the vacuum state undetermined up to the very end and 
to provide a particular Lagrangian depending both on the physical configuration and on the vacuum state.
It is the so-called {\it covariant first order Lagrangian} (see \ref{\Cavalese}, \ref{\Lagrangian}, \ref{\Remarks}, \ref{\Libro}), which is known at the same time to provide a reasonable physical interpretation  for the ambiguity introduced by the vacuum choice, to solve the anomalous factor problem (see \ref{\Kats}, \ref{\ADMCovariante}) and to extend the so-called Regge-Teitelboim prescription to non-asymptotically flat cases in a covariant way.

Still the situation is not completely satisfactory; if the notion of relative conserved quantities has to be introduced as {\it the} fundamental notion of conserved quantities, one still needs to be able to define the analogous of  covariant first order Lagrangians in {\it any} field theory, even when a canonical choice of the vacuum state is available.
Which is what we are going to do in this paper. 
We shall here define an algorithm to define a Lagrangian depending on both a physical field and a vacuum, from which relative conservation laws can be obtained via the usual N\"other theorem through its superpotentials (Gauss-like integrals).
The framework comes out with a satisfactory interpretation and beautiful mathematical properties. Basically these amount to ensure that the vacuum is there without changing the physical contents of the theory (i.e. it does not
change at all the dynamics of the theory); that it is sufficiently general to be selected not to prevent applications of the formalism; and that relative conserved quantities have the expected properties (e.g. the relative energy between two states $A$ and $C$ is the sum of the relative energy between $A$ and $B$ plus
the relative energy between $B$ and $C$ ``whatever'' $B$ is).
Moreover, it agrees with the standard results when a canonical vacuum is available and selected.

\ss
There is another complementary way of seeing the whole framework. Recently many frameworks have been proposed for computing the {\it variation} of conserved quantities (see, \ref{\Silva}, \ref{\Raiteri}, \ref{\Altro} and references quoted therein).
They all rely on some sort of Lagrangian version of the Hamiltonian formulation of the theory (e.g. they all have to do with a Lagrangian formulation of the symplectic form).
There is a large literature available to motivate these frameworks and a lot of applications.
One of the most beautiful applications of them helps to generalize the definition of gravitational entropy of black holes to more general gravitational systems (see \ref{\Taub}). These applications are eventually based on the fact that
the first principle of thermodynamics depends on the variation of conserved quantities while their absolute value is unessential.
All these frameworks define in fact variations of conserved quantities as being of cohomological origin (i.e. independent of any possible divergence addition to the Lagrangian).

Our framework can be seen as a formal integration of all these prescriptions. 
We canonically define a particular representative of the cohomological class of the Lagrangian, which canonically produces relative conservation laws.
These relative conservation laws are the finite form obtained by the integration of the variation of internal energy defined in \ref{\Silva}, \ref{\Raiteri}, \ref{\Altro} along a path in the solution space of the theory.

Let us finally remark that despite our Lagrangians will depend both on the physical fields and on the vacuum state, what we are doing cannot be called a {\it background fixing}. 
The vacuum field, as a matter of facts does not enter in our framework as a {\it background};
of course such a claim depends on what exactly one means by {\it background}, which we stress to be a rather unclear notion in the current literature. 

We claim that what is most often defined as a background in relativistic theories  is an object of the same sort as Minkowski metric one fixes in special relativistic (and particle) Physics (see \ref{\Rovelli}). 
Definitions of this kind imply a number of implicit conventions on how one is allowed  to manipulate the background itself.
It is, for example,  understood that deformations should keep the background fixed when variational calculus is performed.
Hence a background does not obey field equations. For example, in particle Physics the matter Lagrangian (such as Klein-Gordon Lagrangian) depends both on the Minkowski metric and on the matter fields. 
However, when looking for field equations one allows deformations of the matter fields but one keeps the Minkowski metric fixed. 
As a result, field equations (depending on Minkowski metric) are obtained for matter fields, while the Minkowski metric itself has to obey no further equation.
From an equivalent point of view, the Minkowski metric appearing in the matter Lagrangian can be seen as an explicit dependence of the Lagrangian on coordinates (through the Minkowski metric which does in fact depend on coordinates  when general coordinates are adopted for covariance reasons). 
In this context, the matter Lagrangian depends on coordinates as well as on matter fields.

GR is in fact one of the possible solutions to this non-covariant behaviour. If both matter fields and the metric are free to be varied and a free Lagrangian for the metric field interacts with a matter Lagrangian, then the metric is promoted to be a dynamical (and physical) field and a covariant theory is obtained. As a consequence, the metric has to obey its own field equations.
In the variation of the total Lagrangian there are terms  which in the background framework  were constrained to zero by freezing metric variations, and which in the GR framework are not constrained anymore.
These terms do not vanish in general; on the contrary they contribute to metric field equations.
In other words, the absence of backgrounds (in the above sense) is a direct consequence of the principle of general covariance.

Our attitude towards  vacuum fields is perfectly similar to the GR point of view;
deformations act on all fields (vacuum included)  which  have hence to obey their own field equations.

Also background behaviours with respect to symmetries are implicitly prescribed:
in particle physics symmetries (e.g., under the action of  Poincar\'e group) are constrained to preserve the fixed background as well.

This is the source of a consistent misunderstanding in GR framework. In fact, it is  often claimed that particular conservations laws in GR have to be generated by Killing vectors, i.e. by symmetries of the metric. 
It is important to notice that, generically, in GR a solution of Einstein field equations has {\it no} Killing vectors, but luckily enough {\it any} spacetime vector field (not necessarily a Killing vector, if any) is a symmetry for the theory (owing to general covariance) and consequently {\it any} spacetime vector field generates conservation laws, provided N\"other theorem is formulated in the appropriate and correct way.

In other words, particle physics is quite different from GR precisely because of the background fixing.
In particle physics extra terms appear in N\"other theorems, depending on Lie derivatives along symmetry generators acting on the background fields; 
such extra terms have to be constrained to be zero by requiring the symmetries to preserve also the background. Without this assumption, these extra terms prevent in fact the conservation of N\"other currents.
In GR on the contrary,  these extra terms combine together and factorize the metric field equations (which are satisfied on-shell, i.e. along solutions), implying conservation of N\"other currents without any additional requirement.

We stress how the two points analyzed above intertwin among each other: in particle physics one as no field equations for the background since it is considered as being fixed and one has to restrict symmetries to cancel extra terms in conservation laws precisely because the background does not satisfy any equation (i.e. its own field equation).
On the contrary, in GR the metric is promoted to be a dynamical field, and it has to obey its own field equations which provide a general tool to cancel extra terms in conservation laws, without restricting symmetries.

This feature of GR easily extends to all gauge natural theories. 
And once again our attitude towards  vacuum fields is similar to GR:
symmetries drag {\it all} fields and are not constrained to just preserve the vacuum.

A third difference between  vacuum fields and background behaviour resides in their physical interpretation.
Background fields are observable since matter field equations {\it do } explicitely depend on the background.
Particles (or matter fields) move differently if the background is fixed to be an AdS space rather than being Minkowski.
For this reason the background field is endowed with a direct physical meaning.

In GR all observable fields are dynamical. Our vacuum fields {\it do not} break this principle. 
They break instead its logical inverse: vacuum fields are dynamical fields which are not observable.
 As we shall see below, in a theory with vacuum fields, all fields are regarded on an equal footing as far as variations and symmetries are concerned. 
 However, because of the peculiar dynamical structure of the theory, fields decouple in two non-interacting sets, i.e. the physical fields and the vacuum fields (the distinction between the two being to some extent arbitrary on mathematical grounds).
The difference between physical fields and vacuum fields has a {\it physical} origin: owing to the peculiar structure of the interactions the vacuum states are not endowed with a direct physical meaning while the physical fields are.
Vacuum fields do not affect the dynamics of physical fields so that they cannot be observed through interactions.
They just set the zero level of all those additive charges obtained by N\"other conservation laws, as it is reasonable to expect since absolute conserved quantities (e.g. energies) are not endowed with a direct meaning in classical Physics.

We completely agree that from a fundamental point of view physical theories should avoid background fixings, though we believe that vacuum states can be satisfactorily interpreted.

The material in this paper is organized as follows:

\ni 
In  Section $2$ we shall briefly summarize what is known about the variation of conserved quantities.
This Section is introduced with the aim of making the paper self-contained;
for further details we refer to \ref{\Libro}, \ref{\Silva}, \ref{\Raiteri}, \ref{\Altro}.
We shall present the general framework for a generic gauge natural theory. 
This case encompasses all fundamental theories currently used in Physics, as explained in the monograph \ref{\Libro}.

\ni
In Section $3$ we shall define {\it relative conserved quantities} and prove their properties. We shall also 
show how these can be considered as the finite form for the variation of conserved quantities as defined infinitesimally in Section $2$.

\ni
In Section $4$ we shall collect some applications and examples.

\NewSection{Gauge Natural Theories and Variations of Conserved Quantities}

Let us consider a gauge natural theory (see \ref{\Libro}). The structure bundle is fixed to be a principal bundle $P$
with gauge group $G$ over an $m$-dimensional spacetime manifold $M$. 
Natural theories are recovered by just setting $G=\{e\}$.

The configuration bundle $C$ is a gauge natural bundle associated to $P$. 
This amounts to require that there exists a canonical (functorial) action of principal automorphisms of $P$ (also called {\it gauge transformations}) on $C$. 
The framework therefore includes spacetime tensors as well as  tensor densities, spacetime connections, gauge fields, tetrads, spin connections, spinors and all sort of fields currently used in Physics. 
Fibered coordinates on $C$ are denoted by  $(x^\mu, y^i)$. The jet prolongation $J^kC$ of the configuartion bundle is defined to take partial derivatives of fields up to order $k$ into account. 
Natural fibered coordinates on $J^kC$ are denoted by $(x^\mu, y^i, y^i_\mu,\dots, y^i_{\mu_1\dots\mu_k})$,
with an obvious meaning of the notation (see \ref{\Libro} or \ref{\Saunders} for details).

The difference between natural and gauge natural theories is that in natural theories one has spacetime diffeomorphisms acting on fields, while in gauge natural  theories only gauge transformations act.
Vertical gauge transformations (also called {\it pure gauge transformations}) are naturally embedded into the group of gauge transformations while spacetime diffeomorphisms are not. 
In fact, gauge transformations project over spacetime diffeomorphisms while spacetime diffeomorphisms do not embed canonically in gauge transformations.

 Each generator of gauge transformations is a right invariant vector field on $P$, locally in the form
 $$
 \Xi= \xi^\mu(x)\del_\mu+\xi^A(x)\rho_A 
 \fn$$
 where the vector fields $\rho_A$ form a right-invariant pointwise basis for vertical vectors on $P$.
Lie derivatives of fields are defined with respect to $\Xi$ are denoted by $\Lie_\Xi y^i$ (see \ref{\Libro} or \ref{\Trautman} for the general theory and below for applications). 
Lie derivatives of fields are, by construction, linear combinations of  $\xi^\mu$ and $\xi^A$ together with, once suitable connections has been fixed, their symmetrized covariant derivatives up to finite orders (say, $r$ for $\xi^\mu$ and $s$ for $\xi^A$). In natural and gauge natural theories the connections needed to define covariant derivatives are built from dynamical fields (or, as often happens, they {\it are} dynamical fields themselves). 
Hereafter we shall always assume that the covariant derivatives are induced by these {\it dynamical} connections.

A gauge natural Lagrangian $L=\calL(j^k y)\>\d s$ defines the dynamics of the theory.
Here $\d s$ denotes the basis for spacetime $m$-forms ($m$ being the dimension of spacetime); locally
$\d s =\d x^1\land \dots\land \d x^m$ where $m=\dim(M)$.
The Lagrangian is required to be {\it gauge natural}, which amounts to require that gauge transformations are
Lagrangian symmetries.
As a consequence the covariance identity holds (off-shell):
$$
p_i \Lie_\Xi y^i + p_i ^\mu \d_\mu(\Lie_\Xi y^i) +\dots = \d_\mu(\xi^\mu \calL)
\fl{\CovIdEQ}$$
where $(p_i, p_i ^\mu, \dots)$ are the Lagrangian momenta with respect to $(y^i, y^i_\mu,\dots)$, respectively.

By covariant integration of $\CovIdEQ $ by parts with respect to the derivatives of fields' Lie derivatives, 
N\"other theorem is easily proved, giving:
$$
\Div\>\calE(L, \Xi)= \calW(L, \Xi)
\qquad
\cases{
\calE(L, \Xi)= <\F \>| \> j^{k-1} \Lie_\Xi y> - i_\xi  L\cr
\calW(L, \Xi) =- <\E\>|\>  \Lie_\Xi y> \cr
}
\fn$$
Here $\Div$ denotes the formal divergence operator (acting on bundle horizontal forms as 
$(\Div \al)\circ j^{k+1} y= \d( \al\circ j^k y)$ for any section $y$);
$\E: J^{2k}C\arr V^\ast(C)\times A_m(M)$ is called the Euler-Lagrange morphism generating field equations;
$\F: J^{2k-1}C\arr V^\ast(J^{k-1}C)\times A_{m-1}(M)$ is called the Poincar\'e-Cartan morphism (see \ref{\Lagrangian}, \ref{\Poincare} and references quoted therein);
$<\cdot\>|\> \cdot>$ denotes the standard pairing between the bundle of vertical vectors $V(J^kC)$ and its dual bundle $V^\ast(J^kC)\simeq J^kV^\ast(C)$ for the appropriate
value of $k$; $i_\xi$ denotes the contraction of an horizontal form with a spacetime vector field $\xi$.

The Euler-Lagrange morphism is uniquely determined. The Poincar\'e-Cartan morphism is uniquely determined when $k\le 2$, otherwise it depends on a fibered connection to be chosen.
The quantity $\calW(L, \Xi)$ is called {\it work form} and it vanishes on-shell, i.e. along solutions of field equations.
Hence the quantity $\calE(L, \Xi)$, which is called {\it N\"other current}, is conserved (i.e. it is closed) when computed on-shell.

Both quantities $\calW(L, \Xi)$ and $\calE(L, \Xi)$ are horizontal forms on some jet prolongation of the configuration bundle. In view of the fact that Lie derivatives expand as a linear combination of the gauge generator
$\Xi$ and its symmetrized covariant derivatives, both  $\calW(L, \Xi)$ and $\calE(L, \Xi)$ expand as well as linear combinations with horizontal forms as coefficients of the $\Xi$ symmetrized covariant derivatives up to some finite order. These sorts of linear combinations are called {\it variational morphisms} and are related to Spencer cohomology (see \ref{\Libro}, \ref{\Spencer}).

By a suitable, though canonical and algorithmic, covariant integration by parts one obtains then
$$
\calW(L, \Xi)= \calB(L, \Xi) + \Div\> \tilde \calE(L, \Xi)
\qquad
\calE(L, \Xi)= \tilde\calE(L, \Xi) + \Div\>  \calU(L, \Xi)
\fn$$
The quantity $\calB(L, \Xi) $ generates the so-called {\it Bianchi identities} and is identically vanishing  off-shell, i.e. 
on {\it any} configuration not necessarily a solution of field equations;
the quantity $\tilde \calE(L, \Xi)$ is called {\it reduced current} and it is vanishing on-shell;
the quantity $\calU(L, \Xi)$ is usually called {\it superpotential}. The N\"other current, once computed on-shell, is not only conserved (i.e. closed), but also exact regardless of the topology of spacetime and of the configuration bundle; the superpotential is a primitive for it.

All these quantities are uniquely determined when symmetrized covariant derivatives to be integrated by parts
appear just up to second order. For  higher orders the result is a unique function of the fibered connection one uses to define covariant derivatives; in all gauge natural theories one has a fibered connection defined as a function of dynamical fields which is called the {\it dynamical connection}. All these objects are thence uniquely determined once the dynamical connection has been fixed. We stress that one could not single out, e.g., a canonical representative 
for the superpotential by just working with spacetime objects. The canonicity of the representative is obtained just by using the bundle structure of the theory. From now on we shall refer to {\it the} reduced current, superpotential, etc. meaning these canonical representatives. 
Let us also remark that in most cases of physical interest the order involved do not usually exceed two.

 Conserved quantities within a regular region $D$ can be (naively) defined as
 $$
 Q_D(L,\Xi, y)= \int_D \calE(L,\Xi)|_{y}= \int_D \tilde\calE(L,\Xi)|_{y} + \int_{\del D} \calU(L,\Xi)|_{y}
 \fl{\CQEQone}$$
 On the other hand, if we consider the deformation along a family of solutions (generated by some vertical vector field $X$) we obtain
$$
\eqalign{
 \de_X Q_D(L,\Xi, y)=& \int_D \de_X \calE(L,\Xi)|_{y}
 = \int_D \de_X<\F\>|\>\Lie_\Xi>|_{y} - i_\xi <\de L\>|\> X>|_{y}=\cr
 =&\int_D \de_X<\F\>|\>\Lie_\Xi>|_{y} - i_\xi <\E\>|\> X>|_{y} - i_\xi \Div <\F\>|\>X>|_{y}=\cr
 =&\int_D \de_X<\F\>|\>\Lie_\Xi>|_{y} - i_\xi <\E\>|\> X>|_{y}  -\Lie_\xi<\F\>|\>X>|_{y} 
 + \Div( i_\xi<\F\>|\>X>)|_{y}=\cr
 =& \int_D \om(X, \Lie_\Xi y)  - i_\xi <\E\>|\> X>|_{y}
 + \Div( i_\xi<\F\>|\>X>)|_{y}
 \cr}
 \fl{\CQEQtwo}$$
where: $|_y$ means {\it evaluated along a field $y$}; 
we have used the identity $\Lie_\xi = i_\xi\d + \d i_\xi$ holding for operators on spacetime forms; 
we set $\om(X, \Lie_\Xi y)=\de_X<\F\>|\>\Lie_\Xi>_{y}-\Lie_\xi<\F\>|\>X>_{y}$ for the so-called 
{\it symplectic form} (here it is crucial that $\de\Xi=0$; 
if any sort of lift is understood then $\xi^A$ is given as a function of $\xi^\mu$, its derivatives and the dynamical fields, so that some extra integration by parts are needed in that case since $\de_X$ and $\Lie_\Xi$ do not commute; see \ref{\Raiteri}).
The symplectic form defined in this way is exactly analogous to the standard form $\om= \dot p_i\de q^i - \dot q^i \de p_i$ defined in Mechanics see \ref{\Wald}, which using equations of motion coincides with the variation of the Hamiltonian.

Comparing the two expression $\CQEQone$ and $\CQEQtwo$ one obtains
$$
\Div( \de_X\calU(L,\Xi) - i_\xi<\F\>|\>X>)=
\om(X, \Lie_\Xi y)  - i_\xi <\E\>|\> X> + \de_X \tilde\calE(L,\Xi)
\fn$$
Notice that $ i_\xi <\E\>|\> X>$ vanishes on-shell and $\de_X \tilde\calE(L,\Xi)$ vanishes because
$X$ is tangent to the space of solutions.
This suggests to define the {\it corrected conserved quantity density} as
$$
\de_X \calQ(L, \Xi)=\de_X\calU(L,\Xi) - i_\xi<\F\>|\>X>
\fl{\DeQEQ}$$   
We remark that the quantity $\DeQEQ$ is a cohomological invariant; it is in fact independent of any pure divergence term added to the Lagrangian, which does not change Euler-Lagrange field equations but changes the Poincar\'e-Cartan part $\F$ and, consequently, the definition of conserved quantities through $\CQEQtwo$.

There are many motivations to consider seriously this corrected quantity as being fundamental; it overcomes the
anomalous factor problem; it extends Regge-Teitelboim analysis to non-asymptotically flat solutions in a covariant way; it has a nice Hamiltonian meaning; the same expression has been found by different authors relying on different arguments;see,e.g.,  \ref{\Remarks}, \ref{\Silva}, \ref{\Raiteri}, \ref{\Altro}, \ref{\Wald}.

As discussed in \ref{\Raiteri},  if one imposes Dirichlet boundary conditions in a region $D$ (which amounts to require that $\de y^i=0$ on the boundary $\del D$) the integral of  $\de_X \calQ(L, \Xi)$ defines a sort of {\it internal energy} within the region $D$. 
By imposing different boundary conditions (e.g. Neumann conditions, i.e. fixing momenta on the boundary)
other types of energy are recovered (e.g. heat energy).
Hereafter we shall deal with Dirichlet conditions and internal energy. 
Future investigations will be devoted to cover also other cases.

\NewSection{Augmented Lagrangians and Relative Conservation Laws}

We shall take for granted that the variation of conserved quantities in gauge natural theories is given by the
corrected formula $\DeQEQ$.
We shall here prove that we can take advantage of the cohomological invariance of $\DeQEQ$ to
formally integrate it along the family of solutions defined by $X$ obeying Dirichlet boundary conditions.
More formally, we shall prove that, once the vacuum field is introduced, in many cases there is a particular representative (called the {\it augmented Lagrangian}) of the cohomology class of the original Lagrangian for which 
the superpotential provides the formal integration for the variation of conserved quantities.

Let us consider a first order gauge natural Lagrangian $L=\calL(j^1y)$ on $C$; this is not restrictive both because any physical theory has a first order formulation and because it can be easily extended to higher order theories (see below for the case of the second order purely metric GR).
The variation of conserved quantities within $D$ moving from the vacuum state $\bar y$ to the physical state $y=\bar y+\de y$ is given by
$$
\de_X Q_D(L, \Xi, \bar y)=\int_{\del D} \de_X\calQ(L,\Xi)|_{\bar y}
\fn$$

The correction term along a family of solutions generated by $X=(\de y^i) \del_i$ (with $\de y^i=0$ on the boundary of a fixed region $D$) is given by
$$
-i_\xi <\F(\bar L)\>|\> X>= - \bar p_i^{[\mu} \xi^{\nu]} \de y^i \d s_{\mu\nu} 
\fl{\CorEQ}$$
where $\d s_{\mu\nu}$ denotes the ``natural basis'' for spacetime $(m-2)$-forms.

Let us introduce the vacuum state and then consider the augmented Lagrangian on $J^2(C\times_M C)$
$$
l(j^2y, j^2\bar y)= L(j^1 y) - L(j^1\bar y) + \Div\> \al( jy, j\bar y)
\fn$$
where $jy$ and $j\bar y$ denote the jet prolongation to some suitable order.

In general  the superpotential of the augmented Lagrangian is given by
$$
\calU(l, \Xi)= \calU(L, \Xi)- \calU(\bar L, \Xi)+\al^{[\mu}\xi^{\nu]}\d s_{\mu\nu}
\fl{\SupEQ}$$
where we set $\bar L= \calL(j^1 \bar y)\> \d s$.

Let us fix $\al$ so that it obeys the following condition
$$
{\d\over \d s} \al(jy_s, j\bar y)\Big|_{s=0}=-<\F(\bar L)\>|\> X>
\qquad\hbox{with $\al(jy_0, j\bar y)=0$}
\fl{\ConditionEQ}$$
where $y_s$ is the family of solutions generated by integration along $X$ starting from the initial condition $y_0=\bar y$.

\ms\ni{\bf Property: } {\ClaimStyle The superpotential of the augmented Lagrangian $l$ provides a formal integration for the variation of conserved quantities, using the vacuum state as the initial condition.}

\ss\ni{\bf Proof: } {\DimStyle let us compute the generator of the superpotential along the family of solutions generated by $X$:
$$
{\d\over \d s} \calU(l, \Xi)_{(jy_s, j\bar y)}\Big|_{s=0}= \de_X\calU(\bar L,\Xi) + i_\xi {\d\over \d s} \al(jy_s, j\bar y)\Big|_{s=0}= \de_X\calU(\bar L,\Xi) -i_\xi<\F(\bar L)\>|\> X>=\de_X \calQ(\bar L,\Xi)
\fn$$
which proves that $\calU(l, \Xi)$  is the integral of the variation of conserved quantities.\QDE}

\ms\ni
For example, in first order variational principles, by direct comparison with $\CorEQ$, we see that whenever the action of the gauge group on fields  comes as a restriction of a (linear) representation (which encompasses all reasonable cases of fundamental fields, namely metrics, connections, gauge fields, tetrads, tensor densities, spinors, etc.) one can set
$$
\al^\mu(jy_s, j\bar y)= - \bar p_i^\mu (y^i_s-\bar y^i)
\fl{\AlphaEQ}$$
to do the job. We stress that the additional request on the transformation rules of the fields applies just if we pretend that $\al^\mu$ is a global divergence term by itself. If we start from a local Lagrangian $L$ the additional term $\Div \al$ is not expected to have a global meaning either, as for the total augmented Lagrangian $l$ (see below the example about Chern-Simons and non-covariant first order gravitational Lagrangian).

\ms\ni{\bf Property: } {\ClaimStyle The condition $\ConditionEQ$ has at most one essential solution for $\al$.}

\ss\ni{\bf Proof: }  {\DimStyle If there were two solutions, $\al$ and $\be$, their difference would obey:
$$
{\d\over \d s}(\al-\be)\Big|_{s=0}=0
\fn$$
along all families of solutions. Hence $\al-\be$ is the extension of a function which is constant on the space of solutions.
The global constant can be fixed by requiring that $\al=\be$ at $s=0$ (i.e. by fixing the initial value for $\bar y$).

Since we shall be interested with the value of $\al$ on the solution space only, we can say that it is essentially unique.\QDE
}

\ms
We remark that the existence of the quantity $\al$ is not exactly a well posed problem. Of course a local integration
for $\al$ can always and easily be produced by simply expanding any infinitesimal quantity $\de y$ into a finite difference $y-\bar y$.
This naive solutions has to pass at least two tests to be accepted: 
first, the finite difference appearing in this way should be a covariant object (here the exact meaning of covariant depends on the context);
second, the augmented Lagrangian defined by the choice of $\al$ must be covariant or gauge covariant depending on the context.
Further future investigations will be devoted to the problem of characterizing the existence of the integral quantity $\al$ in fully general settings.

\NewSection{Applications and Examples}

In the previous Section we have reduced the problem of formal integration of conserved quantity to the solution of
$\ConditionEQ$. Hereafter we shall show a number of examples which show that the condition $\ConditionEQ$
can be actually solved for many theories of current interest in Physics.

\NewSubSection{Purely metric Hilbert Lagrangian}

Consider the Lagrangian $L_H= R\sqrt{g}\>\d s$ thought as a second order Lagrangian for a metric $g$.
Its Poincar\'e-Cartan morphism is (see \ref{\Lagrangian})
$$
<\F\>|\> j^1X>= \sqrt{g} \>g^{\al\be}\de u^\la_{(\al\be)}\>\d s_\la
\fn$$
where $u^\la_{(\al\be)}= \Ga^\la_{\al\be} -\de^\la_{(\al} \Ga_{\be)}$, $\Ga^\la_{\al\be}$ denotes the coefficients of the Levi-Civita connection (namely Christoffel symbols) and $\Ga_{\be}= \Ga^\al_{\al\be}$
denotes the trace of the connection. Here $\d s_\la$ is the local basis of spacetime $(m-1)$-forms.

Let us hence define $w^\la_{(\al\be)}= u^\la_{(\al\be)}- \bar u^\la_{(\al\be)}$ and the augmented Lagrangian
$$
l(jg, j\bar g)= R\sqrt{g}- \bar R\sqrt{\bar g} -d_\la ( \sqrt{\bar g}\> \bar g^{\al\be}w^\la_{(\al\be)})
\fn$$ 
This is known in the literature as the covariant first order Lagrangian for $g$ and $\bar g$ (being first order in $g$).
It generalizes Regge-Teitelboim presciption to non-asymptotically flat cases (asymptotically flat cases are recovered by setting $\bar g$ equal to the Minkowski metric).
There is a whole bunch of papers motivating its use for computing conserved quantities (see, e.g.,  \ref{\Cavalese}, \ref{\Lagrangian}, \ref{\Libro}, \ref{\Taub}, \ref{\BTZ}, \ref{\HDTaub} and references quoted therein).

Let us remark that the Poincar\'e-Cartan morphism for the augmented Lagrangian can be computed as
$$
\eqalign{
<\F(l)\>|\> j^1 X>=& \left[\sqrt{g} g^{\al\be} \de u^\la_{(\al\be)}
-\sqrt{\bar g} \bar g^{\al\be} \de \bar u^\la_{(\al\be)}
-\de(\sqrt{\bar g}\> \bar g^{\al\be})w^\la_{(\al\be)}
-\sqrt{\bar g}\> \bar g^{\al\be}\de u^\la_{(\al\be)}
+\sqrt{\bar g}\> \bar g^{\al\be}\de \bar u^\la_{(\al\be)}
\right]\>\d s_\la=\cr
=&\left[(\sqrt{g} g^{\al\be} - \sqrt{\bar g}\> \bar g^{\al\be})\de u^\la_{(\al\be)}
-\de(\sqrt{\bar g}\> \bar g^{\al\be})w^\la_{(\al\be)}
\right]\>\d s_\la\cr
}
\fn$$
which vanishes once Dirichlet boundary conditions (namely $\de g^{\al\be}=0$ as well as $g^{\al\be}=\bar g^{\al\be}$) are imposed.

\NewSubSection{Palatini Lagrangian}

Let us consider the so-called Palatini formalism for GR (also known as first order formalism). 
Fundamental fields are a metric $g$ and a torsionless connection $\Ga$.
The Lagrangian is defined as 
$$
L= \sqrt{g}\> g^{\al\be} R_{(\al\be)}\> \d s
\fn$$
where $R_{(\al\be)}$ denotes the symmetric part of the Ricci tensor of $\Ga$.

The Poincar\'e-Cartan morphism is 
$$
<\F\>|\> X>= \sqrt{g}\> g^{\al\be}\de u^\la_{\al\be}\>\d s_\la
\fn$$
where $u^\la_{\al\be}= \Ga^\la_{\al\be} -\de^\la_{\al} \Ga_{\be}$ and $\Ga_{\be}= \Ga^\al_{\al\be}$
denotes the trace of the connection.

Let us hence define $w^\la_{\al\be}= u^\la_{\al\be}- \bar u^\la_{\al\be}$ and the augmented Lagrangian by
$$
l(g, \Ga ,\bar g, \bar \Ga)= R_{(\al\be)}g^{\al\be}\sqrt{g}- \bar R_{(\al\be)}\bar g^{\al\be}\sqrt{\bar g} 
-d_\la ( \sqrt{\bar g} \>\bar g^{\al\be}w^\la_{\al\be})
\fn$$ 
which is simply the covariant first order Lagrangian in Palatini formalism.

\NewSubSection{Non-covariant first order Einstein Lagrangian}

Let us consider the {\it non-covariant first order Einstein Lagrangian}. It is obtained by subtracting out the second
derivatives of the metrics  $g$ and $\bar g$ under a divergence. Of course this procedure produces a local Lagrangian depending on the coordinate patch, since the discarded divergence cannot be global if spacetime has non-trivial topology.
The Lagrangian is
$$
L = \left( \sqrt{g}\> g^{\al \be} R_{\al \be} - d_\la (\sqrt{g}\> g^{\al \be}  u^\la_{\al\be})\right) \>\d s
\fn$$
This Lagrangian is known to produce the same equations as second order Hilbert Lagrangian, both in purely metric and in Palatini framework.

The Poincar\'e-Cartan morphism is given by
$$
\eqalign{
<\F\>|\>j^1 X>=& \left(  \sqrt{g}\> g^{\al \be}  \de u^\la_{\al\be}-\de(\sqrt{g}\> g^{\al \be}  u^\la_{\al\be})\right)\>\d s_\la=\cr
=& -\left(  \de(\sqrt{g}\> g^{\al \be})  u^\la_{\al\be}\right)\> \d s_\la=\cr
\cr}
\fn$$
so that we set
$$
\al^\la=(\sqrt{ g}\> g^{\al \be} - \sqrt{\bar g}\> \bar g^{\al \be}) \bar  u^\la_{\al\be}
\fn$$
The augmented Lagrangian is hence given by
$$
\eqalign{
l=& \sqrt{g}\> g^{\al \be} R_{\al \be}- \sqrt{\bar g}\> \bar g^{\al \be} \bar R_{\al \be}
+\d_\la (\sqrt{\bar g}\> \bar g^{\al \be}  \bar u^\la_{\al\be} - \sqrt{g}\> g^{\al \be}  u^\la_{\al\be}
+\sqrt{g}\> g^{\al \be}\bar  u^\la_{\al\be} - \sqrt{\bar g}\> \bar g^{\al \be}\bar  u^\la_{\al\be}) =\cr
=&\sqrt{g}\> g^{\al \be} R_{\al \be}- \sqrt{\bar g}\> \bar g^{\al \be} \bar R_{\al \be}
-\d_\la ( \sqrt{g}\> g^{\al \be}  (u^\la_{\al\be} -\bar  u^\la_{\al\be}))=\cr
=&\sqrt{g}\> g^{\al \be} R_{\al \be}- \sqrt{\bar g} \bar g^{\al \be} \bar R_{\al \be}
-\na_\la ( \sqrt{g}\> g^{\al \be}  w^\la_{\al\be} )=
\sqrt{g}\> g^{\al \be} R_{\al \be}- \sqrt{\bar g} \bar g^{\al \be} \bar R_{\al \be}
-\bar\na_\la ( \sqrt{g}\> g^{\al \be}  w^\la_{\al\be} )
\cr}
\fn$$
which, despite we started from a different representative in cohomology class (or better a {\it local} one which
has not even the right to be called a Lagrangian), we still reproduce the same result already obtained for the classically equivalent Hilbert (or Palatini, as can be easily proved) Lagrangian.
This shows how the notion of augmented Lagrangian is robust.

\NewSubSection{Chern-Simons}

Let us now consider Chern-Simons Lagrangian. For simplicity let us fix the group to be $G=\SO(m)$ and denote its elements by $S^i_j$; it is understood that these parameters, which are actually coordinates on $\GL(m)$,
are constrained by the relations $^tS S=\one$ and $\det\>S=1$.

Let $P$ be a principal $G$-bundle over $M$; a right invariant  pointwise basis for vertical vectors is
given by $\si_{ij}=\rho_{[ij]}$ where we set
$$
\rho_{ij} = \de_{hi}S^{h}_k\>{\del\over\del S^j_k}
\fn$$
A principal connection is given by
$$
\om= \d x^\mu\otimes(\del_\mu-A^{ij}_{\mu}(x)\si_{ij})
\fn$$
Let us also set $A^i_\mu=\ep^i{}_{jk} A^{jk}_\mu$.
Chern-Simons Lagrangian is a local Lagrangian 
(which provides global field equations, as it happens also for the non-covariant first order Einstein Lagrangian);
on a $3$-dimensional base $M$, and with obvious notation, it is  given by
$$
L= \ep^{\al\be\la}\left(\eta_{ij} F^i{}_{\al\be} A^j{}_{\la} - {1\over 3}\ep_{ijk} A^i{}_{\al}A^j{}_{\be}A^k{}_{\la}\right)
\fn$$
We remark that Lie algebra indices are raised and lowered by $\de_{ij}$.

The Poincar\'e-Cartan morphism is given by (see \ref{\NostroCS}):
$$
<\F\>|\> X>=2\ep^{\al\be\la} \eta_{ij}\de A^i{}_{\be} A^j{}_{\la} \> \d s_\al
\fn$$
which defines the following correction term
$$
\al^\mu=-2\ep^{\mu\be\la} \eta_{ij}( A^i{}_{\be} - \bar A^i{}_{\be}) \bar A^j_{\la}
\fn$$
This correction is non-covariant, but we stress that the original Lagrangian was non-covariant either;
what has to be covariant is the final augmented Lagrangian while the single terms can be non-covariant.
We also remark that the contribution of $\ep^{\mu\be\la}  \eta_{ij}\bar A^i{}_{\be} \bar A^j{}_{\la}$ vanishes due to symmetry properties.
The augmented Lagrangian is given by
$$
\eqalign{
l=&  \ep^{\al\be\la}\left(\eta_{ij} F^i{}_{\al\be} A^j{}_{\la} - {1\over 3} \ep_{ijk}A^i{}_{\al}A^j{}_{\be}A^k_{\la}\right)
- \ep^{\al\be\la}\left(\eta_{ij} \bar F^i_{\al\be} \bar A^j{}_{\la} - {1\over 3} 
\ep_{ijk}\bar A^i{}_{\al}\bar A^j{}_{\be}\bar A^k{}_{\la}\right)+\cr
&-\d_\mu(2\ep^{\mu\be\la}  \eta_{ij}A^i{}_{\be} \bar A^j{}_{\la})=\cr
=& \ep^{\mu\nu\rho}\left(
\eta_{ij} \bar F^i_{\mu\nu} B^j_\rho +
\eta_{ij} \bar \na_\mu B^i_{\nu} B^j_\rho
+{1\over 3} \ep_{ijk}B^i{}_{\al}B^j{}_{\be}B^k_{\la}
\right)\cr
}
\fn$$
where we set $B^i_\mu=A^i_\mu-\bar A^i_\mu$ and $\bar \na_\mu$ denotes the covariant derivative induced by the vacuum connection $\bar A^i_\mu$.  

The second expression given for the Lagrangian $l$ shows that it is in fact gauge covariant.
This Lagrangian is already known as the {\it Chern-Simons covariant Lagrangian}; see \ref{\NostroCS}.

\NewSubSection{Higher order gravitational Lagrangians}

Let us consider a Lagrangian which is a generic function of the Ricci scalar $\calL=\sqrt{g}f(R)$.
The Poincar\'e-Cartan morphism reads as
$$
<\F\>|\> X>=( \sqrt{g} f'(R)  g^{\al\be} \de u^\mu_{\al\be}) \>\d s_\mu
\fn$$
The correction term is given by
$$
\al^\mu= \sqrt{\bar g} f'(\bar R)  \bar g^{\al\be} w^\mu_{\al\be}
\fn$$
and the augmented Lagrangian is given by
$$
l=\sqrt{g}f(R)- \sqrt{\bar g}f(\bar R) +\na_\mu(\sqrt{\bar g} f'(\bar R)  \bar g^{\al\be} w^\mu_{\al\be}) 
\fn$$

\ms\ni
Let us now consider a Lagrangian which is a generic function of the Ricci tensor squared, i.e. $\calL=\sqrt{g}f(R_{\mu\nu}R^{\mu\nu})$.
The Poincar\'e-Cartan morphism reads now as
$$
<\F\>|\> X>=(2\sqrt{g} f'(R_{\rho\si}R^{\rho\si}) R^{\al\be} \de u^\mu_{\al\be})\>\d s_\mu
\fn$$
where we set $u^\mu_{\al\be}= \Ga^\mu_{\al\be} - \de^\mu_\al \Ga_\be$ to acknowledge the possibility that Ricci tensor may be non-symmetric.
The correction term is given by
$$
\al^\mu=-2\sqrt{\bar g} f'(\bar R_{\rho\si}\bar R^{\rho\si}) \bar R^{\al\be} w^\mu_{\al\be}
\fn$$
and the augmented Lagrangian is given by
$$
l=\sqrt{g}f(R_{\mu\nu}R^{\mu\nu})- \sqrt{\bar g}f(\bar R_{\mu\nu}\bar R^{\mu\nu}) 
-2\na_\mu(\sqrt{\bar g} f'(\bar R_{\rho\si}\bar R^{\rho\si}) \bar R^{\al\be} w^\mu_{\al\be}) 
\fn$$

\ms\ni
Let us finally consider a Lagrangian which is a generic function of the Riemann tensor  squared, i.e. $\calL=\sqrt{g}f(R_{\al\be\mu\nu}R^{\al\be\mu\nu})$
The Poincar\'e-Cartan morphism reads in this case as
$$
<\F\>|\> X>=4(\sqrt{g} f'(R_{\al\be\mu\nu}R^{\al\be\mu\nu}) R_\al{}^{\be\mu\nu} 
\de \Ga^\al_{\be\nu})\>\d s_\mu
\fn$$
The correction term is given by
$$
\al^\mu=-4\sqrt{\bar g} f'(\bar R_{\ep\la\rho\si}\bar R^{\ep\la\rho\si}) \bar R_\al{}^{\be\mu\nu} 
 (\Ga^\al_{\be\nu}- \bar \Ga^\al_{\be\nu})
\fn$$
and the augmented Lagrangian is given by
$$
l=\sqrt{g}f(R_{\al\be\mu\nu}R^{\al\be\mu\nu})
- \sqrt{g}f(\bar R_{\al\be\mu\nu}\bar R^{\al\be\mu\nu}) 
-4\na_\mu(\sqrt{\bar g} f'(\bar R_{\ep\la\rho\si}\bar R^{\ep\la\rho\si}) \bar R_\al{}^{\be\mu\nu} 
 q^\al_{\be\nu}) 
\fn$$
where we set $q^\al_{\be\nu}= \Ga^\al_{\be\nu}- \bar \Ga^\al_{\be\nu}$.

We remark that many higher order Lagrangians considered in the literature, e.g. the effective Lagrangians arising from low energy limit of string theory as well as Lovelock theories (see \ref{\Lovelock}), are just combinations of these basic examples.

\NewSubSection{Yang-Mills Lagrangian}

Let us consider the Yang-Mills Lagrangian $\calL= -{\sqrt{g}\over 4} F^A_{\mu\nu} F_A^{\mu\nu}$
(possibly coupled with a gravitational Lagrangian) for any structure Lie group $G$.
The Poincar\'e-Cartan morphism is 
$$ 
<\F\>|\> X>=-{\sqrt{g}} F_A^{\mu\nu} \de  A^A_{\nu}
\fn$$
Hence the correction term is
$$
\al^\mu=  {\sqrt{\bar g}} \bar F_A^{\mu\nu} (A^A_{\nu}-\bar A^A_{\nu})
\fn$$
The augmented Lagrangian is hence
$$
l=  -{\sqrt{g}\over 4} F^A_{\mu\nu} F_A^{\mu\nu} +{\sqrt{\bar g}\over 4} \bar F^A_{\mu\nu} \bar F_A^{\mu\nu}
+\na_\mu \left({\sqrt{\bar g}} \bar F_A^{\mu\nu} (A^A_{\nu}-\bar A^A_{\nu})\right)
\fn$$

We stress that the correction introduced to the conserved quantities is proportional to $\bar F_A^{\mu\nu}$
and hence vanishes whenever a flat vacuum state is chosen.

Let us also mention that the Poincar\'e-Cartan morphism of the augmented Lagrangian is
$$
\eqalign{
<\F(l)\>|\> X>=& \left[\sqrt{g} F_A^{\mu\nu} \de A^A_\nu
-\sqrt{\bar g} \bar F_A^{\mu\nu} \de \bar A^A_\nu
+\de(\sqrt{g} \bar F_A^{\mu\nu}(A^A_{\nu}-\bar A^A_{\nu}))
\right]\>\d s_\mu=\cr
=&\left[-(\sqrt{g} F_A^{\mu\nu} -\sqrt{\bar g} \bar F_A^{\mu\nu})\de A^A_\nu
+\de(\sqrt{\bar g} \bar F_A^{\mu\nu})(A^A_{\nu}-\bar A^A_{\nu})
\right]\>\d s_\mu\cr
}
\fn$$
which vanishes once Dirichlet conditions ($\de A^A_\mu=0$ as well as $A^A_\mu=\bar A^A_\mu$) are imposed.

\NewSection{Conclusions and Perspectives}

The technique we discussed can be interesting for dealing with the inverse problem of variational calculus.
If fact, we showed as augmented variational principles can be used to globalize local variational principles
(see Chern-Simons and non-covariant Einstein first order Lagrangian).

If compared with previous papers of some of us about the covariant first order Lagrangian, some apparent inconsistencies can be found.
Usually we used the correction term as $\tilde \al^\mu = -\sqrt{g}\> g^{\al} w^\mu_{\al\be}$ instead of
$\al^\mu = -\sqrt{\bar g}\>\bar g^{\al} w^\mu_{\al\be}$ which ensues from the algorithmic procedure.
We stress, however, that in those papers we checked conserved quantities by integrating at spatial infinity where
appropriate boundary conditions are imposed (namely, $g=\bar g$). Hence no difference in conserved quantities is found when using $\tilde\al$ in place of the canonical $\al$, since the two choices are in a sense equivalent once boundary conditions have been imposed.

In connection with this, let us moreover remark that when boundary conditions are imposed there is always an ambiguity in the surface terms since there one has $y=\bar y$. We guess however that the difference is always inessential to physically relevant quantities.

Future investigations will be devoted to boundary conditions different from Dirichlet which are known to generate different ``thermodynamical'' energies for the system.
Another issue to be addressed in future investigations is the behaviour of the augmented Lagrangians under Legendre transformations.

 \NewSection{Appendix A}

One can easily check that the relative energy between a material point of mass $m$ held steady at height $h$
in a constant gravitational field and the same point steady at height $\bar h=0$ is $E=mg(h-\bar h)$.
This is independent of the (inertial) observer; if the whole system is on a train travelling with uniform velocity $w$
the computation of $E$ is not affected by $w$.
One could (erroneously) argue that this happens because,  in that case,  the weight force is orthogonal to the train displacement so that the work done by the force is zero.
This argument is erroneous for a number of reasons. We hereafter present a different example where the force
and the displacement are not orthogonal, but the relative energy is still independent of the observer.
 
Let us consider two material points $(P_1, m)$ and $(P_2, m)$ linked by an ideal spring (with a spring constant $k$).
The whole system is contrained along a straight line (e.g. the $x$-axis) and it is on a train moving 
with constant velocity  $w$ along that $x$-axis.
The initial conditions are set so that the points initially sit at the equilibrium position with opposite initial velocities.
The evolution of the system is given by
$$
\cases{
x_1= wt + A\cos(\om t)\cr
x_2= wt - A\cos(\om t)\cr
}
\qquad
\om^2=2 k^2/m
\fn$$
Hence the system, seen from the train, is simply oscillating in a harmonic fashion. 
The energy of the system is 
$$
\eqalign{
E=& {m\over 2}\left( w^2 + A^2\om^2\sin^2(\om t) +2wA\om \sin(\om t)\right)
+ {m\over 2}\left( w^2 + A^2\om^2\sin^2(\om t) -2wA\om \sin(\om t)\right)+\cr
&+ {k^2\over 2} 4 A^2 \cos^2(\om t)=
m w^2 + mA^2\om^2\sin^2(\om t)  + 2{k^2} A^2 \cos^2(\om t)
=  m w^2 + m A^2 \om^2 \cr
}
\fn$$
Of course, this {\it energy} is conserved along the motion, but being the energy with respect to the observer frame,
it depends on the observer through $w$.

Now suppose one considers two solutions, corresponding to the amplitudes $A_1$ and $A_2$ respectively.
One can compute the corresponding energies $E_1$ and $E_2$.
Both these energies are conserved and both depend on the observer through $w$.

Now the difference $E_2-E_1$ is conserved {\it and} it does not depend on the reference frame any longer, being:
$$
E_2-E_1=  m (A_2^2-A_1^2) \om^2
\fn$$
This depends just on the  two configurations and it is covariant (i.e. independent of the observer).

\ni [If a simple harmonic oscillator is considered, the work done against the constraint---which fixes one side of the spring to the ground--- messes up the things a bit. Luckily enough this sort of constraints do not exist in fundamental field theories to which we want to extend this stuff. Let us stress that a simple harmonic oscillator is precisely obtained by the system here considered by a sort of background fixing $x_2=wt$. Hence we could say that the possibility of defining relative conserved quantities is directly related exactly to the absence of backgrounds in the theory, i.e. with the fact that vacuum state is dynamical in our framework.]

We have so dealt with on-shell quantity defined along a class of solutions. The same result can be obtained through 
variational techniques which are independent of the knowledge of any explicit solution.
\ms
\centerline {\Molla}
\ss
\centerline {Fig. 1 --- Lagrangian coordinates}
\ms
When the Lagrangian coordinates are choosen as in Figure $1$,
the system above is described by the Lagrangian
$$
L= {m\over 2}\left(v_1^2+ v_2^2\right)-{k^2\over 2 } \left(x_1-x_2\right)^2
\fn$$
By changing to baricentrical coordinates $x={x_1+x_2\over 2}$, $q={x_1-x_2\over 2}$ the Lagrangian is recast in the form:
$$
L= {m}\left(w^2+ u^2\right)- 2k^2 q^2
\fn$$
where $w=\dot x$ and $u=\dot q$ denote Lagrangian velocities.
The corresponding energy with respect to the observer is obtained by computing along a solution the expression
$$
\ep= {m}\left(w^2+ u^2\right)+ 2k^2 q^2
\fn$$

Let us now consider the Lagrangian
$$
\eqalign{
\tilde L =& L-\bar L =
\left({m}\left(w^2+ u^2\right)- 2k^2 q^2\right)
-\left({m}\left(\bar w^2+ \bar u^2\right)- 2k^2 \bar q^2\right)\cr
}
\fn$$
It depends on both a dynamical configuration $(x, q)$ and a vacuum state $(\bar x, \bar q)$ to be chosen
in the space of solutions.
If N\"other theorem is applied the energy of this  augmented Lagrangian is easily obtained
$$
\eqalign{
\ep-\bar \ep =&
\left({m}\left(w^2+ u^2\right)+ 2k^2 q^2\right)
-\left({m}\left(\bar w^2+ \bar u^2\right)+ 2k^2 \bar q^2\right)=\cr
=&{m} \left(w^2-\bar w^2\right) + 
{m\over 2}\left(u^2- \bar u^2\right)+ 2k^2 \left(q^2-\bar q^2\right)
\cr
 }
\fn$$ 
which differs from the {\it relative energy} $E$ computed above along a particular solution by the term
$\De E={m}(w^2- \bar w^2)$.
This difference is zero whenever the two solutions, i.e. both the dynamical section and the vacuum,
have the same baricentrical velocity ($w=\bar w$).
We can hence restrict the two solutions to be two different oscillating modes on the same train  (i.e. $w=\bar w$);
with these additional boundary conditions we defined a perfect prescription for the computation of 
the {\it relative energy}.
It is also easy to prove that $E$ is {\it covariant} (i.e. independent of the inertial observer), namely invariant with respect to to transformations
$$
(x',q', \bar x', \bar q')=(x+ wt,q, \bar x+ wt, \bar q)
\qquad \forall w\in \R
\fn$$
The reader might try to generalize as an exercise this example to harmonic oscillators in uniform motion one with respect to the other. For us the above example will suffice as a mechanical metaphoric introduction to the general field theory case.

More generally if a (discrete or continuum) system  of material points is considered and just internal forces are present, K\"onig theorem shows that total energy splits into the {\it baricentrical kinetic energy} and the {\it relative mechanical energy} (see, e.g., \ref{\Goldstein}). Of course, baricentrical kinetic energy is observer dependent but independent of the internal configuration while the relative mechanical energy is observer independent.
The difference between the energies of two different configurations (with the same baricentrical speed) is observer independent.

Our proposal is to use this sort of {\it relative conserved quantities} as the only conserved quantities endowed 
with a fundamental physical meaning, since they obey covariance principle {\it a priori}.
The observer dependent conserved quantities are nevertheless important when one has to deal with the Newtonian (or post-Newtonian) limit of a general relativistic model since the quantities we are used to define in Newtonian Physics have historically developed in a non-relativistic framework. However, {\it relative conserved quantities} are the only fundamental quantities which are available in a pure relativistic context.
The price we are forced to pay is the {\it relative} character of them.
We do not have much choice: either we accept the {\it relative} character of these quantities (i.e. the dependence on two configurations) or we accept their dependence on the observer (i.e. non-covariance).

\NewSection{References}

\Biblio

\end